\documentclass[12pt,a4paper]{article}
\usepackage{amsmath}
\usepackage{amssymb}
\usepackage[nosort]{cite}
\usepackage{hyperref}
\usepackage{graphicx}
\usepackage{color}
\usepackage{float}
\usepackage{mathrsfs}
\allowdisplaybreaks

\makeatletter
\@addtoreset{equation}{section}

\makeatother

\makeatletter
\let\old@makecaption=\@makecaption
\def\@makecaption{\small\old@makecaption}
\makeatother

\makeatletter
\let\old@startsection=\@startsection
\renewcommand{\@startsection}[6]{\old@startsection{#1}{#2}{#3}{#4}{#5}{#6\mathversion{bold}}}
\makeatother

\let\oldPhi=\Phi
\let\oldPsi=\Psi
\let\oldGamma=\Gamma
\let\oldSigma=\Sigma
\renewcommand{\Phi}{\mathnormal{\oldPhi}}
\renewcommand{\Psi}{\mathnormal{\oldPsi}}
\renewcommand{\Gamma}{\mathnormal{\oldGamma}}
\renewcommand{\Sigma}{\mathnormal{\oldSigma}}

\newcommand{\hypref}[2]{\ifx\href\asklfhas #2\else\href{#1}{#2}\fi}

\newcommand{\half}{\frac{1}{2}}

\newcommand{\nn}{\nonumber}

\newenvironment{myeqnarray}{\arraycolsep0pt\begin{eqnarray}}{\end{eqnarray}\ignorespacesafterend}
\newenvironment{myeqnarray*}{\arraycolsep0pt\begin{eqnarray*}}{\end{eqnarray*}\ignorespacesafterend}

\def\[{\begin{equation}}
\def\]{\end{equation}}
\def\<{\begin{myeqnarray}}
\def\>{\end{myeqnarray}}

\newcommand{\qe}{\begin{eqnarray}}
\newcommand{\ee}{\end{eqnarray}}
\newcommand{\na}{\nabla}
\newcommand{\esp}{\, \, \, \,}
\newcommand{\es}{\, \,}
\newcommand{\esum}{\,}
\newcommand{\la}{\lambda}
\newcommand{\qua}{\frac{1}{4}}

\ifx\href\asklfhas\newcommand{\href}[2]{#2}\fi

\begin{document}

\thispagestyle{empty}

\vspace{1cm}

\renewcommand{\thefootnote}{\fnsymbol{footnote}}
\setcounter{footnote}{0}

\begin{center}

{\Large\textbf{\mathversion{bold}On the Pure Spinor Heterotic Superstring  $b$ Ghost}\par}

\vspace{1cm}

\textsc{Thiago Fleury}
\vspace{8mm}

\textit{
Instituto de F\'isica Te\'orica, UNESP - Univ. Estadual Paulista, \\
ICTP South American Institute for Fundamental Research, \\
Rua Dr. Bento Teobaldo Ferraz 271, 01140-070, S\~ao Paulo, SP, Brasil} 
\vspace{2mm}

\textit{and}

\vspace{2mm}

\textit{Perimeter Institute for Theoretical Physics, \\
Waterloo, Ontario N2L 2Y5, Canada}

\vspace{8mm}

\texttt{tfleury@ift.unesp.br}\par\vspace{1cm}

\textbf{Abstract}\vspace{5mm}

\begin{minipage}{13.7cm}

A simplified pure spinor superstring $b$ ghost in a curved heterotic background was constructed recently. The $b$ ghost is a composite operator and it is not holomorphic. However, it satisfies $\bar\partial b = [ Q , \Omega ]$, where $Q$ is the BRST charge. In this paper, we find a possible $\Omega$.

\end{minipage}

\end{center}

\newpage
\setcounter{page}{1}
\renewcommand{\thefootnote}{\arabic{footnote}}
\setcounter{footnote}{0}

\tableofcontents
\section{Introduction}

The minimal pure spinor  superstring action in a generic supergravity and super-Yang-Mills (SYM)  heterotic background first appeared in \cite{BerkovitsHowe}. 
It was also shown in \cite{BerkovitsHowe} that imposing that both the BRST operator is nilpotent and the BRST current is holomorphic all the supergravity and super-Yang-Mills superfield constraints are reproduced. However, the $b$ ghost in a curved heterotic background was only fully constructed recently \cite{Heteroticb}.    

In the pure spinor formalism, the $b$ ghost is a composite operator and it is constructed as a solution to the equation $\delta_B b =T$, where $\delta_B$ means BRST variation and $T$ is the energy-momentum tensor.  In order to construct a covariant $b$ ghost in a flat space background, it is necessary to use the non-minimal pure spinor formalism introduced in \cite{Topological}. The left-moving ghost sector of the minimal formalism consist of a pure spinor $\la^{\alpha}$ and its conjugate momentum $\omega_{\alpha}$. The non-minimal formalism has the following additional variables: a bosonic pure spinor $\hat\la_{\alpha}$ and a constrained fermion $r_{\alpha}$ together with their conjugate momenta $\hat\omega^{\alpha}$ and $s^{\alpha}$ respectively. In \cite{RNSlike}, the $b$ ghost in a flat space background was simplified, it was rewritten in terms of RNS-like variables.   

The BRST transformations of all the minimal fields in a curved heterotic background were computed in \cite{BRSTminimal} and they were used to show that 
the heterotic minimal action is BRST invariant. The non-minimal formalism in a curved heterotic background was constructed in \cite{ChandiaNonMinimal}. The same set of non-minimal variables described previously for the flat space background was added to the model and their BRST transformations were found requiring that the BRST transformations of the minimal variables are unchanged and imposing some consistency conditions. This non-minimal heterotic formalism was used to construct the heterotic $b$ ghost. A simplified version of the curved heterotic $b$ ghost, and the one used in this work, was constructed in \cite{Heteroticb}.             

In a flat space background $b$ is an holomorphic operator, i.e. $\bar\partial b =0$ where $\bar\partial$ is the antiholomorphic derivative. However, it is not holomorphic in a generic curved background
and it is expected to satisfy $\bar\partial b = \delta_B \Omega$ for some $\Omega$. Note that $\Omega$ is defined up to a BRST exact term. In fact, $\Omega$ has been constructed in a Type IIB  $AdS^5 \times S^5$ background in \cite{Taming} and in a super-Maxwell background in \cite{Maxwell}. In this work, we find a possible  $\Omega$ for the curved heterotic $b$ ghost of \cite{Heteroticb}.   

The pure spinor $b$ ghost in a generic curved Type II background has not been constructed yet. It is only known for particular R-R backgrounds \cite{Taming}, being $AdS^5 \times S^5$ an example.  
Recently, the non-minimal pure spinor formalism in a generic curved Type II background was constructed in \cite{NonMinimalFormalism}. It is expected that the $b$ ghost can be constructed using this new non-minimal formalism.

This paper is organized as follows. In section 2, we review both the minimal and non-minimal pure spinor formalisms in a curved heterotic background. Moreover, the expression of the $b$ ghost is presented and all the necessary equations of motion to compute $\bar\partial b$ are derived. In section 3, we explicitly construct $\Omega$ firstly in a super-Yang-Mills heterotic background, then in the case  $s^{\alpha}=\hat\omega^{\alpha}=0$ and finally the general case is considered. The complete expression for $\Omega$ is given in section 4. The Appendix has all the conventions used in this work and some important identities.

\section{Review of the Pure Spinor Formalism in a Curved Heterotic Background} 

In this section, we review both the minimal and non-minimal pure spinor heterotic superstring. Moreover, the expression of the $b$-ghost is presented. We also derive all the necessary equations of motion to evaluate $\bar\partial b$. All the conventions used can be found in the Appendix. 

\subsection{The Minimal Pure Spinor Formalism}

The minimal pure spinor heterotic superstring action $S_m$  is written in terms of the $\mathcal{N}=1$ $D=10$ superspace coordinates and background superfields.
We will denote the curved superspace coordinates by $Z^M$ with $M=(m, \mu)$ and $m=0, \ldots ,9$, $\mu=1, \ldots, 16$. The tangent superspace indices will be denoted by $A=(a, \alpha)$, with $a=0, \ldots ,9$, and $\alpha = 1, \ldots, 16$.  The action in a generic curved heterotic background is \cite{BerkovitsHowe}  
\begin{eqnarray}
S_{m}= \int d^2z  \;  \frac{1}{2} \Pi^a \overline{\Pi}^b \eta_{ab} + \frac{1}{2} \Pi^{A} \overline{\Pi}^{B} B_{BA} + d_{\alpha} \overline{\Pi}^{\alpha} + w_{\alpha} \overline{\na} \lambda^{\alpha}  \label{eq:theactionminimal} \hspace{15mm} 
 \\ 
+ J_I (\Pi^A A_{A}^I + d_{\alpha} W^{I \alpha} + \frac{1}{2} \lambda^{\alpha} w_{\beta}  U^{I  \beta} _{\alpha}) + S_J  + \bar{b} \partial \bar{c}  \, , \nn 
\end{eqnarray} 
with
\begin{eqnarray}
\Pi^A = \partial Z^M E_M^{\esp \es A} \, , \quad  \quad  \, \,  \overline{\Pi}^{A} = \overline{\partial} Z^M E_M^{\esp \es A} \, , \nn 
\end{eqnarray}
where $E_{M}^{\esp \es A}$  is  the supervielbein matrix. In the action, $J_I$ are right-moving $E_8 \times E_8$ or $SO(32)$ currents with $I=1, \ldots, 496$ and their action is $S_J$, whose explicit form will not be needed. The $(\bar{b}, \bar{c})$ are the usual right-moving Virasoro ghosts of the bosonic string. The additional ghosts are $( \la^{\alpha} , \omega_{\alpha} )$ with $\omega_{\alpha}$ being the conjugate momentum of $\la^{\alpha}$ and $\la^{\alpha}$ a bosonic pure spinor, which means that it satisfies 
\begin{eqnarray}
\la \gamma_a \la = 0 \, ,  \nn 
\end{eqnarray} 
where $(\gamma_a)_{\alpha \beta}$ are tangent $SO(1,9)$ generalized Pauli matrices. The covariant derivative that appears in the action is given by
\begin{eqnarray}
\overline{\nabla} \lambda^{\alpha} = \overline{\partial} \lambda^{\alpha} + \lambda^{\beta} \overline{\Pi}^A \Omega_{A \beta}^{\quad \alpha} \, , \nn 
\end{eqnarray} 
with $\Omega_{M \beta}^{\esp \esp \es \alpha}$ the background spin connection. The additional background superfields appearing in the action are:
the two-form potential $B_{MN}$, the super-Yang-Mills potential $A_M^I$ and finally, the super-Yang-Mills field-strengths $W^{I \alpha}$ and $U^{I \beta}_{\alpha}$.  

Due to the fact that $\la^{\alpha}$ is a constrained variable, its momentum conjugate $\omega_{\alpha}$ is defined up to the gauge transformation
\begin{eqnarray}
\delta \omega_{\alpha}= \Lambda^{a} (\gamma_a \la)_{\alpha} \, ,  
\label{eq:gaugeomega}
\end{eqnarray}
for any $\Lambda^a$. Imposing that the action is gauge invariant implies that the superfields $\Omega_{M \alpha}^{\esp\esp\es \beta}$ and $U^{I \beta}_{\alpha}$ 
have the decompositions
\begin{eqnarray}
\Omega_{A \alpha}^{\esp \esp \es \beta} = \Omega_A \, \delta^{\beta}_{\alpha} + \frac{1}{4} \Omega_{A ab} (\gamma^{ab})_{\alpha}^{\esp \beta} \, , \quad \quad U^{I \beta}_{\alpha} = U^I \delta^{\beta}_{\alpha} + \frac{1}{4} U^I_{ab} (\gamma^{ab})_{\alpha}^{\esp \beta} \, , \label{eq:DecompositionOU}
\end{eqnarray}
which is equivalent to the fact that $\omega_{\alpha}$ can only appears in the gauge invariant combinations $J= \la^{\alpha} \omega_{\alpha}$ and $N^{ab}=\half (\la \gamma^{ab} \omega)$.   
In the decomposition of the spin connection above, $\Omega_A$ is the scaling connection and $\Omega_{A ab}$ is the usual Lorentz connection satisfying  
$\Omega_{M ab} = - \Omega_{M ba}$.

The action $S_m$ is BRST invariant and the left-moving BRST operator is 
\begin{eqnarray}
Q = \oint dz  \, \la^{\alpha} d_{\alpha}  \, .
\end{eqnarray} 
Note that the variable $d_{\alpha}$ also appears in the action. The pure spinor BRST operator was first constructed in a flat space background in \cite{BerkovitsFirst} and its origin, at least for the flat space case, was only discovered recently as a result of the procedure of gauge-fixing a more fundamental action \cite{BerkovitsSecond,BerkovitsThird}. 

The pure spinor formalism is well defined only if the BRST operator is nilpotent and the BRST current is holomorphic, which means $\bar\partial (\la d) =0$. In \cite{BerkovitsHowe} the constraints on the background superfields coming from these two conditions were found and solved and it was shown that they are the correct 
$\mathcal{N}=1$ supergravity and super-Yang-Mills constraints after fixing all the gauge symmetries. In this work, we will need the following set of these constraints         
\begin{eqnarray}
T_{A \beta}^{\esp \esp \alpha} = H_{b a \beta} = T_{\beta (ab)} =0 \, , \quad H_{\alpha \beta b}=T_{\alpha \beta b}\, ,  \label{eq:Constraints} \hspace{25mm}  \\ \nn \\
\quad \la^{\beta} \la^{\gamma} R_{b \beta \gamma}^{\esp \esp \esp \alpha} =0 \, , \quad \, \, F_{\alpha \beta}^I =0 \, ,  \quad \, \, F_{a \beta}^I = W^{I \alpha} (\gamma_a)_{ \alpha \beta} \, , \quad  \half U^{I \alpha}_{\beta}= \nabla_{\beta} W^{I \alpha}  \, ,  \nn \\ \nn \\
T_{\alpha \beta}^{\esp \esp a} = \gamma^a_{\alpha \beta} \, , \quad T_{\alpha a}^{\esp \esp b} = 2 (\gamma_{a}^{\es b})_{\alpha}^{\esp \beta} \Omega_{\beta} \, , \hspace{32mm}  \nn
\end{eqnarray}
where $H$ are the three-form field-strengths, $T$ are the torsions, $R$ are the curvatures, $F$ are the SYM field-strengths and $\nabla_M$ are covariant derivatives defined in terms of the spin connections and SYM connections, see the Appendix for precise definitions.  

In fact, the action $S_m$ of ($\ref{eq:theactionminimal}$) is not the complete action. The Fradkin-Tseytlin term that couples the background dilaton superfield $\Phi$ with the worldsheet curvature is absent. This term was not written down because it is higher order in $\alpha'$ and it is not needed for classical conformal invariance of the action. However, it is necessary for quantum conformal invariance \cite{BerkovitsHowe,ConventionsCV} and homorphicity of the BRST current $(\la d)$, which imply the additional constraint
\begin{eqnarray}
\Omega_{\alpha} = \qua \nabla_{\alpha} \Phi \, .
\end{eqnarray}

Combining the set of constraints given above with the Bianchi identities, one can derive additional relations among the background superfields. Using that  
$\nabla_A \gamma^b_{\alpha \beta} = - 2 \Omega_A \gamma^{b}_{\alpha \beta}$ and the Bianchi identity involving the super-Yang-Mills field-strength $F^I_{AB}$ given in
(\ref{eq:BianchiF}), one has
\qe
\nabla_{[a} F^I_{\alpha \beta]} + T_{[a \alpha}^{\esp \esp F} F^I_{|F| \beta]} =0 \esp \Rightarrow   \quad \half U^I + \Omega_{\alpha} W^{I \alpha} = 0\, \, \,  {\rm{and}} \, \,  F^I_{ab}= \half U^I_{ab} + T_{\alpha ab} W^{I \alpha} \, . \label{eq:RelationsU}
\ee

Moreover, three important properties of the background superfields can be derived from a combination of two Bianchi identities. Consider the Bianchi identities (\ref{eq:BianchiT}) and (\ref{eq:BianchiH}) with the following specification of the indices   
\begin{eqnarray}
\eta^{ab} (\nabla_{[a} T_{\alpha \beta] b} - R_{[a \alpha \beta] b} + T_{[a \alpha}^{ \esp \esp  A} T_{|A| \beta] b}) =0 \, , \esp {\rm{and}} \esp 
\nabla_{[a} H_{b \alpha \beta]} + \frac{3}{2} T_{[a b}^{\esp \esp A} H_{|A| \alpha \beta]} =0 \, ,  \nn
\end{eqnarray}
one can show that the Bianchi identities above imply 
\begin{eqnarray}
\Omega_a = 0 \, , \esp \esp  \esp \eta^{ab}T_{c a b} = 0 \, , \esp \esp \esp T_{abc} + H_{abc}=0 \, . \label{eq:TwithH}
\end{eqnarray}

In what follows, the explicit form of $R_{AB}$, which are the components of the curvature constructed from the scaling connection $\Omega_{A}$, will be needed. Note that for generic values of $A$ and $B$, the curvature has the form      
\qe
R_{AB} = \nabla_{[A} \Omega_{B]} + T_{AB}^{\esp \esp C} \Omega_C \, , \nn  
\ee
and due to the constraints above, we have 
\qe
R_{a \beta} = \nabla_a \Omega_{\beta} \, , \esp \quad R_{ab} = T_{ab}^{\esp \es \alpha} \Omega_{\alpha} \, , \esp \quad  R_{\alpha \beta} = \nabla_{(\alpha} \Omega_{\beta)} \es \es  \Rightarrow \es \es R_{\alpha \beta} = - \qua \gamma^a_{\alpha \beta} \nabla_a \Phi \, . \label{eq:TheRTwoIndices}
\ee
\subsubsection{The BRST transformations} \label{BRST}
As mentioned before, the minimal action $S_m$ of (\ref{eq:theactionminimal}) is BRST invariant. The BRST transformations of all the variables appearing on the minimal action were derived in \cite{BRSTminimal} expressing $d_{\alpha}$ as a function of the other variables and the conjugate momenta and using the canonical commutation relations. The transformations of $\la^{\alpha}$, $\omega_{\alpha}$, $d_{\alpha}$ and $J_I$ are 
\begin{eqnarray}
\delta_B \lambda^{\alpha} = - \lambda^{\beta}  \Omega_{\beta \gamma}^{\esp \esp  \alpha} \lambda^{\gamma} \, , \quad \es \delta_B \omega_{\alpha} = \lambda^{\beta} \Omega_{\beta \alpha}^{\esp \esp \gamma}\omega_{\gamma}  + d_{\alpha} \, , \quad \es \delta_{B} J_I = J_K  f^{K}_{\esp J I} \, \lambda^{\alpha} A_{\alpha}^J \, ,
\\ \nn \\
\delta_B d_{\alpha} = \la^{\beta} \Omega_{\beta \alpha}^{\esp \esp \gamma} d_{\gamma} + \la^{\beta} \la^{\gamma} \omega_{\delta} R_{\alpha \beta \gamma}^{\esp \esp \esp \delta} + \Pi^a (\la \gamma_a)_{\alpha} \, . \hspace{20mm}  \nn
\end{eqnarray}
In addition, the BRST transformations of $\Pi^A$ are  
\begin{eqnarray}
\delta_{B} \Pi^a = - \la^{\alpha}  \Omega_{\alpha b}^{\esp \esp a} \Pi^b - \la^{\alpha} \Pi^A T_{A \alpha}^{\esp \esp a} \, , \quad \es  \delta_B \Pi^{\beta} = \nabla \la^{\beta} - \la^{\alpha} \Omega_{\alpha \gamma}^{\esp \esp \beta} \Pi^{\gamma} \, .
\end{eqnarray}
The transformations of the variables $\overline{\Pi}^A$ can be obtained from the ones above by replacing both $\Pi$ with $\overline\Pi$ and $\partial$ with $\bar\partial$. Finally, let $\Psi$ be any superfield, it transforms as
\qe
\delta_B \Psi = \lambda^{\alpha} \partial_{\alpha} \Psi \, . 
\ee


In the expression of the heterotic $b$ ghost derived in \cite{Heteroticb} which will be reviewed in a future subsection, the following combination of variables and
superfields appears   
\qe
D_{\alpha} = d_{\alpha} + \frac{1}{4} \la^{\beta} T_{\beta a b} (\gamma^{ab})_{\alpha}^{\esp \gamma} \omega_{\gamma} - 3 (\la \Omega) \omega_{\alpha}  \, .
\label{eq:DefinitionD}
\ee
The BRST transformation of $D_{\alpha}$ was derived in \cite{Heteroticb}. It was shown that using the constraints of the superfields and the Bianchi identities, the transformation simplifies to 
\qe
\delta_B D_{\alpha} = \la^{\beta} \Omega_{\beta \alpha}^{\esp \esp \gamma} D_{\gamma} + 3 (\la \Omega) D_{\alpha} + \Pi^a (\la \gamma_a)_{\alpha} - \qua \la^{\beta} T_{\beta ab} (\gamma^{ab} D)_{\alpha} \, . 
\ee
One last comment is that sometimes the computations are simplified once one writes the BRST transformation of $\omega_{\alpha}$ in terms of $D_{\alpha}$, it takes the form
\qe
\delta_{B} \omega_{\alpha} = \la^{\beta} \Omega_{\beta \alpha}^{\esp \esp \gamma} \omega_{\gamma} + D_{\alpha} + 3 (\la \Omega) \omega_{\alpha} - \frac{1}{4} \la^{\beta} T_{\beta ab} (\gamma^{ab})_{\alpha}^{\esp \gamma} \omega_{\gamma}  \, .
\ee

\subsection{The Non-Minimal Pure Spinor Formalism}

The non-minimal pure spinor formalism in a flat background was developed in \cite{Topological}. 
One of the motivations for adding new variables to the minimal formalism was that using these new variables one can construct a covariant $b$ ghost which is important for
multi-loop calculations. The new variables are: a bosonic pure spinor $\hat\la_{\alpha}$ together with its conjugate momentum $\hat\omega^{\alpha}$ and a constrained fermionic spinor $r_{\alpha}$ and its conjugate momentum  $s^{\alpha}$. They satisfy       
\qe
\hat\la \gamma_a \hat\la=0 \, , \quad \esp  \esp \esp  \hat\la \gamma_{a} r =0 \, , 
\ee
and due to the above constraints the conjugate momenta are defined up to the gauge transformations  
\qe
\delta \hat\omega^{\alpha} = \hat\Lambda^a (\gamma_a \hat\la)^{\alpha} - \phi^a (\gamma_a r)^{\alpha} \, , \quad \esp \esp \delta s^{\alpha} = \phi^a (\gamma_a \hat\la)^{\alpha} \, , \nn
\ee
where $\hat\Lambda^a$ and $\phi^a$ are arbitrary parameters. 

The non-minimal pure spinor formalism in a curved heterotic background was constructed in \cite{ChandiaNonMinimal}. Again, one of the motivations for extending the minimal formalism was the construction of the $b$ ghost and the same set of new variables was added.  
The first step of the construction was finding the BRST transformations of the non-minimal variables after modifying the BRST operator by adding a new term equal to the flat space case. 
This was achieved by noticing that the transformations of the minimal fields do not change and imposing consistency of the BRST transformations. The answer is 
\qe
\delta_B \hat\la_{\alpha} = - r_{\alpha} + \lambda^{\gamma} \hat\la_{\beta}(\Omega_{\gamma \alpha}^{\esp \esp \beta} - \frac{1}{4} T_{\gamma ab}(\gamma^{ab})_{\alpha}^{\esp \beta}) \, , \esp   \esp
\delta_B \hat\omega^{\alpha} = - \hat\omega^{\beta} \lambda^{\gamma}(\Omega_{\gamma \beta}^{\esp \esp \alpha} - \frac{1}{4} T_{\gamma ab}(\gamma^{ab})_{\beta}^{\esp \alpha})  \, , \nn \\ \nn \\
\delta_B s^{\alpha} = \hat\omega^{\alpha} + s^{\beta} \la^{\gamma}(\Omega_{\gamma \beta}^{\esp \esp \alpha}-\frac{1}{4} T_{\gamma ab} (\gamma^{ab})_{\beta}^{\esp \alpha}) \, , \esp \esp \es 
\delta_B r_{\alpha} = - \la^{\gamma} r_{\beta} (\Omega_{\gamma \alpha}^{\esp \esp \beta} - \frac{1}{4} T_{\gamma ab} (\gamma^{ab})_{\alpha}^{\esp \beta}) \, . 
\nn \\ \label{eq:BRSTnon}
\ee

The non-minimal pure spinor action that we are going to use includes a torsion term as the one in \cite{Heteroticb}. The action is   
\qe
S_{nm} = \delta_B \int d^2 z ( s \overline{\nabla} \hat\la + \frac{1}{4} \overline{\Pi}^A T_{A ab} (s \gamma^{ab} \hat\la)) \, , \nn 
\ee
with 
\qe
\overline{\nabla} \hat\lambda_{\alpha} = \bar{\partial} \hat\lambda_{\alpha} - \hat\lambda_{\beta} \overline{\Pi}^A \Omega_{A \alpha}^{\esp \esp \beta} \, ,  \nn 
\ee
and after computing the BRST transformations, one has 
\begin{eqnarray}
S_{nm} = \int d^2 z  \esp \hat\omega \overline{\nabla} \hat\la + s \overline{\nabla} r + \frac{1}{4} \overline{\Pi}^A T_{A ab} [ (\hat{w} \gamma^{ab} \hat\la) + (s \gamma^{ab} r)]  \label{eq:NonMinimalAction} \\
+ \la^{\alpha} \overline{\Pi}^A R_{A \alpha} (s \hat\la) 
+ \frac{1}{4} \la^{\alpha} \overline{\Pi}^c  \mathcal{R}_{c \alpha ab}  (s \gamma^{ab} \hat\la) \, ,  \hspace{10mm} \nn 
\end{eqnarray} 
and we have used the definition
\qe
\mathcal{R}_{c \alpha ab} = R_{c \alpha ab} - \nabla_{[c}T_{\alpha] ab} - T_{c \alpha}^{\esp \esp d} T_{d a b} + T_{c d [ a} T_{b] \alpha}^{\esp \esp d}  \, ,
\label{eq:DefinitionRT}
\ee
where $[ \es ] $ means antisymmetrization without any additional numerical factor. Note that replacing $c$ by $\beta$ in the above definition gives $\mathcal{R}_{\beta \alpha ab} =0$ due to the Bianchi identity (\ref{eq:BianchiT}).   
 

\subsection{The Curved Heterotic $b$ ghost} 
In the pure spinor formalism the $b$ ghost is a composite operator and it is constructed as a solution to the equation $\delta_B b = T$ where $T$ is the energy-momentum tensor. In a flat space background this equation was solved in \cite{Topological} using the non-minimal formalism. The result was rewritten using RNS-like vector variables  in a simpler way in \cite{RNSlike}. The $b$ ghost in a curved heterotic background was first constructed in \cite{ChandiaNonMinimal}, however in this paper we will use the simplified version of the $b$ ghost of \cite{Heteroticb} where generalized RNS-like variables were used. The $b$ ghost is given by
\qe
b = - s^{\alpha} \nabla \hat\la_{\alpha} - \frac{1}{4} \Pi^A T_{A ab} (s \gamma^{ab} \hat\la) - \omega_{\alpha} \Pi^{\alpha} \hspace{45mm} \label{eq:bghost} \\ +\frac{1}{2 ( \la \hat\la)} (\omega \gamma_a \hat\la)(\la \gamma^{a} \Pi) + \Pi^a \bar{\Gamma}_{a} - \frac{1}{4 ( \la \hat\la)} (\la \gamma^{ab} r) \bar{\Gamma}_a \bar{\Gamma}_b \, , \nn 
\ee   
where $\bar\Gamma_a$ are the generalized RNS-like variables defined by
\qe
\bar{\Gamma}_{a} = - \frac{1}{2( \la \hat\la)} (D \gamma_{a} \hat\la) - \frac{1}{8 (\la \hat\la)^2} (r \gamma_{abc} \hat\la) N^{bc} \, , \label{eq:GeneralizedRNS}
\ee
with $D_{\alpha}$ defined in (\ref{eq:DefinitionD}) and $N^{bc}=\half (\la \gamma^{bc} \omega)$.

\subsection{The Equations of Motion}

In this subsection, we compute all the necessary equations of motion for evaluating $\bar\partial b$. The complete curved heterotic action is the sum of the minimal action $S_m$ of  (\ref{eq:theactionminimal}) and the non-minimal action $S_{nm}$ of (\ref{eq:NonMinimalAction}). Varying $\la^{\alpha}$, $\omega_{\alpha}$ and $d_{\alpha}$, one obtains respectively    
\qe
\bar{\partial} \omega_{\alpha} = \overline{\Pi}^A \Omega_{A \alpha}^{\esp \esp  \beta} \omega_{\beta}  +  \half  J_I  U^{I \beta}_{\alpha} \omega_{\beta}  + \overline{\Pi}^A R_{A \alpha} (s \hat\la) + \frac{1}{4} \overline{\Pi}^c \mathcal{R}_{c \alpha ab} (s \gamma^{ab} \hat\la) \, , \\ \nn \\
\bar{\partial} \la^{\alpha} = - \la^{\beta} \overline{\Pi}^A \Omega_{A \beta}^{\esp \esp \alpha} - \half \la^{\beta}  J_I  U^{I \alpha}_{\beta} \, , \quad \quad
\overline{\Pi}^{\alpha} = - J_I W^{I \alpha} \, ,  \hspace{12mm} \nn 
\ee
where $\mathcal{R}_{c \alpha ab}$ was defined in (\ref{eq:DefinitionRT}). 

One way of obtaining the equation of motion of $J_I$ is writing $S_J$ in  (\ref{eq:theactionminimal}) explicitly and varying the action with respect to the right-moving variables.  Another way, which is independent of any specific representation of $J_I$, is to note that $J_I$ transforms under gauge transformations and when $W^{I \alpha}=U^{I \beta}_{\alpha}=0$, its equation of motion has to be $\nabla J_I=0$. Inspecting the action $S_m$ of (\ref{eq:theactionminimal}), one verifies that $W^{I \alpha}$ and $U^{I \beta}_{\alpha}$ couple to $J_I$ similarly to $A_M^I$, then when these background superfields are non-zero the equation of motion of $J_I$ is                   
\qe
\partial J_I = - f^{K}_{\esp I J} J_K (\Pi^A A_A^J + d_{\alpha} W^{J \alpha} + \half \la^{\alpha} \omega_{\beta} U^{J \beta}_{\alpha}) \, .
\ee

 The procedure to find the equation of motion of $d_{\alpha}$ is to first vary the action with respect to $Z^M$ and then multiply the result by the inverse supervielbein $E_{\alpha}^{\esp M}$. Firstly, let us set $\hat\omega^{\alpha}=s^{\alpha}=0$ which is equivalent to consider only the minimal action. In this case, one can show that    
\begin{eqnarray}
\bar{\partial} d_{\alpha}|_{s=\hat\omega=0} = - \half \Pi^A \overline{\Pi}^B H_{BA \alpha} + \half  \Pi^a \overline{\Pi}^{A} T_{A \alpha a} + \half \overline{\Pi}^a \Pi^A T_{A \alpha a} - \overline{\Pi}^A T_{A \alpha}^{\esp \esp \beta} d_{\beta} 
+ \overline{\Pi}^A \Omega_{A \alpha}^{\esp \esp \beta} d_{\beta} \nn \\ \nn \\ + \la^{\beta}  \omega_{\gamma} \overline{\Pi}^A R_{A \alpha \beta}^{\esp \esp \esp \gamma} +  \Pi^A J_I F_{A \alpha}^I + J_I  \nabla_{\alpha} W^{I \beta} d_{\beta} - \half  \la^{\beta} \omega_{\gamma} J_I \nabla_{\alpha} U^{I \gamma}_{\beta} \, . \hspace{15mm} \nn 
\end{eqnarray}
The expression above greatly simplifies when the constraints of the superfields are taken into account. The result is 
\qe
\bar{\partial} d_{\alpha}|_{s=\hat\omega=0} = \overline{\Pi}^{A} \Omega_{A \alpha}^{\esp \esp \beta} d_{\beta} +  \la^{\beta} \omega_{\gamma} \overline{\Pi}^A R_{A \alpha \beta}^{\esp \esp \esp \gamma} - \half \la^{\beta} \omega_{\gamma} J_I \nabla_{\alpha} U^{I \gamma}_{\beta} + \half J_I U^{I \beta}_{\alpha} d_{\beta} \,. \nn 
\ee
Adding the terms coming from the variation of the non-minimal action and using the definition of $\mathcal{R}_{c \beta ab}$ given in (\ref{eq:DefinitionRT}), the final result is  
\qe
&\bar{\partial} d_{\alpha} = \overline{\Pi}^{A} \Omega_{A \alpha}^{\esp \esp \beta} d_{\beta} + \la^{\beta} \omega_{\gamma}  \, \overline{\Pi}^A R_{A \alpha \beta}^{\esp \esp \esp \gamma} - \half \la^{\beta} \omega_{\gamma} J_I \nabla_{\alpha} U^{I \gamma}_{\beta} + \half J_I U^{I \beta}_{\alpha} d_{\beta}  \label{eq:eomd} 
\\ \nn \\ 
&- \half \la^{\beta} J_I U^{I \gamma}_{\beta} R_{\gamma \alpha} (s \hat\la) 
- \overline{\Pi}^A R_{A \alpha} [(sr)+(\hat\omega \hat\la)] + \la^{\beta} \overline{\Pi}^A \nabla_{\alpha} R_{ \beta A} (s \hat\la) + \la^{\beta} \overline{\Pi}^A \nabla_A R_{\alpha \beta} (s \hat\la)  \nn  \\ \nn \\
&+ \la^{\beta} \overline{\Pi}^A T_{A \alpha}^{\esp \esp a} R_{a \beta} (s \hat\la) 
- \qua \la^{\beta} \overline{\Pi}^c \nabla_{\alpha} \mathcal{R}_{c \beta a b} (s \gamma^{ab} \hat\la) 
- \qua \overline{\Pi}^c \mathcal{R}_{c \alpha ab} [(s \gamma^{ab} r)+(\hat\omega \gamma^{ab} \hat\la)] 
\nn \\  \nn \\ &+ \half \la^{\beta} \overline{\Pi}^c \mathcal{R}_{c \beta a d} T_{\alpha b}^{\esp \esp d} (s \gamma^{ab} \hat\la) + \qua \la^{\beta} \overline{\Pi}^A T_{A \alpha}^{\esp \esp c} \mathcal{R}_{c \beta ab} (s \gamma^{ab} \hat\la)  \, . 
\nn
\ee

In what follows, we will also need  $\bar\partial \Pi^{\alpha}$ and $\bar\partial \Pi^a$. One way of computing these equations of motion is using the definitions of $\Pi^A$ and $\overline\Pi^A$ and evaluating sums and differences. Note that
\qe  
(\bar\partial \Pi^A - \partial \overline{\Pi}^A) = \partial Z^M \bar\partial Z^N  \partial_{[N} E_{M]}^{\esp \es  A} = \hspace{3mm} \label{eq:Inter} \\ \nn \\
- \overline{\Pi}^B \Pi^C T_{CB}^{\esp \esp A} - \Pi^B \overline{\Pi}^C \Omega_{CB}^{\esp \esp A} + \overline{\Pi}^B \Pi^C \Omega_{CB}^{\esp \esp A} \, , \nn 
\ee
where we have used the expression for the torsion (\ref{eq:TorsionDef}). Using the expression above and $\overline{\Pi}^{\alpha}= - J_I W^{I \alpha}$, one finds one of the required equations of motion
\qe
\bar{\partial} \Pi^{\alpha} = \partial (- J_I W^{I \alpha} ) - \overline{\Pi}^a \Pi^b T_{ba}^{\esp \es \alpha} - \Pi^{\beta} \overline{\Pi}^C \Omega_{C \beta}^{\esp \esp \alpha} + \overline{\Pi}^{\beta} \Pi^C \Omega_{C \beta}^{\esp \esp \alpha} \, .   
\ee

In order to find the equation of motion $\bar\partial \Pi^a$, it is necessary to use (\ref{eq:Inter}) with $A$ replaced by $a$ and additionally compute    
\qe
\bar\partial \Pi^a + \partial \overline{\Pi}^a = 2 \, \bar\partial \partial Z^M E_{M}^{\esp a} + \partial Z^M \bar\partial Z^N \partial_N E_M^{\esp \es a} + \bar\partial Z^M \partial Z^N \partial_N E_M^{\esp \es a}  \, . \label{eq:Inter22}
\ee
The combination of terms appearing on the right-hand side of the expression above also appears when varying the action with respect to $Z^M$ and multiplying the result by $E_{a}^{\esp  M}$. In this way, one can compute the left-hand side.  Combining (\ref{eq:Inter}) with (\ref{eq:Inter22}), we have


\qe
\bar\partial \Pi_a = \Pi^b \overline{\Pi}^A \Omega_{A a b} + \overline{\Pi}^b T_{b a}^{\esp \es \alpha} d_{\alpha} + \la^{\alpha} \omega_{\beta} \overline{\Pi}^B R_{a B \alpha}^{\esp \esp \esp \beta}   - J_I \nabla_a W^{I \alpha} d_{\alpha} + \half \Pi^b J_I U^{I}_{a b}   \hspace{5mm}   \label{eq:eomPi} \\ \nn \\
+ \half \la^{\alpha} \omega_{\beta} J_I \nabla_a U^{I \beta}_{\alpha} + \overline{\Pi}^A R_{A a}[(sr)+(\hat\omega \hat\la)] + \la^{\alpha} \overline{\Pi}^A \nabla_a R_{A \alpha} (s \hat\la) + \la^{\alpha} \overline{\Pi}^A \nabla_A R_{\alpha a} (s \hat\la)  \nn \\ \nn \\ - \la^{\alpha} \overline{\Pi}^A T_{A a}^{\esp \es B} R_{B \alpha} (s \hat\la) 
+ \half \la^{\alpha} J_I U^{I \beta}_{\alpha} R_{a \beta} (s \hat\la) - \qua \overline{\Pi}^A \mathfrak{R}_{a A b c} [(s \gamma^{bc} r)+(\hat\omega \gamma^{bc} \hat\la)]  \hspace{6mm} \nn \\ \nn \\ + \qua \la^{\alpha} \overline{\Pi}^b \nabla_a \mathcal{R}_{b \alpha c d} (s \gamma^{cd} \hat\la)  
- \qua \la^{\alpha} \overline{\Pi}^A \nabla_A \mathcal{R}_{a \alpha b c} (s \gamma^{bc} \hat\la) + \half \la^{\alpha} \overline{\Pi}^b \mathcal{R}_{b \alpha c d} T_{a e}^{\esp \es d} (s \gamma^{ce} \hat\la) \hspace{2mm}  \nn \\ \nn \\ - \qua \la^{\alpha} \overline{\Pi}^A T_{A a}^{\esp \es b} \mathcal{R}_{b \alpha c d} (s \gamma^{cd} \hat\la) 
- \half \la^{\alpha} \overline{\Pi}^A T_{A b}^{\esp \es c} \mathcal{R}_{a \alpha c d}(s \gamma^{bd} \hat\la) + \frac{1}{8} \la^{\alpha} J_I U^{I \beta}_{\alpha} \mathcal{R}_{a \beta bc} (s \gamma^{bc} \hat\la) \, ,  \nn
\ee
where we have used the definition
\qe
\mathfrak{R}_{c A ab} = R_{c A ab} + \nabla_{[A} T_{c] ab} + T_{A c}^{\esp \es B} T_{B ab} - T_{A d [ a} T_{b ] c}^{\esp \es d} \, ,     \label{eq:DefinitionRTT}
\ee
and (\ref{eq:DefinitionRT}). Note that when $A$ is replaced by $\alpha$ in the above expression, one has 
$\mathfrak{R}_{c \alpha ab} = \mathcal{R}_{c \alpha ab}$. In what follows, we will also need the equations of motion of the non-minimal variables. Varying the action with respect to $\hat\omega^{\alpha}$, $r_{\alpha}$ and $s_{\alpha}$ respectively, we have     
\qe
\bar{\partial} \hat{\la}_{\alpha} =  \overline{\Pi}^A \Omega_{A \alpha}^{\esp \esp  \beta} \hat{\la}_{\beta} - \frac{1}{4} \overline{\Pi}^A T_{A ab} (\gamma^{ab} \hat{\la})_{\alpha} \, , \quad \esp  
\bar{\partial} s^{\alpha} = - s^{\beta} \overline{\Pi}^A \Omega_{A \beta}^{\esp \esp \alpha} + \frac{1}{4} \overline{\Pi}^A T_{A ab} (s \gamma^{ab})^{\alpha} \, ,  \nn \\ \nn \\
\bar{\partial} r_{\alpha} = \overline{\Pi}^{A} \Omega_{A \alpha}^{\esp \esp \beta}  r_{\beta} - \frac{1}{4} \overline{\Pi}^A T_{A ab} (\gamma^{ab} r)_{\alpha} + \la^{\beta} \overline{\Pi}^A R_{A \beta} \hat\la_{\alpha} + \frac{1}{4} \la^{\beta} \overline{\Pi}^d \mathcal{R}_{d \beta ab} (\gamma^{ab} \hat\la)_{\alpha} \, .  \hspace{5mm}
\ee

\section{Constructing $\Omega$}

In this section, we find a $\Omega$ satisfying $\bar\partial b = \delta_B \Omega$. Firstly, the case where all the supergravity background superfields are set to zero will be considered. Secondly, we will consider the case where $\hat\omega^{\alpha}=s^{\alpha}=0$. Finally, the general case will be considered.      

\subsection{Super-Yang-Mills background} 

We begin this subsection by reviewing the construction of $\Omega$ in a super-Maxwell background (open superstring) done in \cite{Maxwell}. The authors of \cite{Maxwell} argued  that $\bar\partial b$ can be extracted from the OPE between $b$ and the super-Maxwell vertex operator. In the limit where the field-strengths are constant, the super-Maxwell vertex operator reduces to a linear combination of the supersymmetry current and the minimal Lorentz current. As the $b$ ghost is supersymmetric and a Lorentz scalar, they concluded that in that limit $\Omega$ is    
\qe
\qua \hat\la_{\beta} (\gamma^{ab})_{\alpha}^{\esp \beta} F_{ab} \frac{\partial b}{ \partial r_{\alpha}}   \, . 
\ee

In this subsection, returning to our construction of $\Omega$ in a curved heterotic background, the case where all supergravity background superfields are set to zero will be considered. In this case, the $b$ ghost expression given in (\ref{eq:bghost}) reduces to the flat space one, because $D_{\alpha} \rightarrow d_{\alpha}$. In addition, our linearized equations of motion take the same functional form of the equations of motion of an open superstring in a generic super-Maxwell background of \cite{Maxwell}, apart from the overall $\delta(\sigma)$ term and including $J_I$ and the indices $I$ on the superfields. Note that, due to (\ref{eq:RelationsU}), we have
$U^{I \beta}_{\alpha} \rightarrow \half (\gamma^{ab})_{\alpha}^{\esp \beta} F^I_{ab}$. Thus, in the case $\partial_a W^{I \alpha}= \partial_a F^I_{bc}=0$, or in other words considering the field-strengths constants, we can use the results of \cite{Maxwell} and conclude that 
\qe
\Omega_0  = \frac{1}{8} (J_I U^I_{ab} ) \, \hat\la_{\beta} (\gamma^{ab})_{\alpha}^{\esp \beta} \, \frac{\partial b}{\partial r_{\alpha}} \, , \label{eq:ConstantLi}
\ee
where the subscript 0 in $\Omega_0$ means that $\Omega$ reduces to this expression when all the supergravity background superfields are zero, the SYM field-strengths are constant and the equations of motion are linearized. In principle, we can use the results of \cite{Maxwell} for the case where the field-strengths are not constant as well, however in \cite{Maxwell}  the $b$ ghost is not written in terms of RNS-like variables, thus we will derive it again below. 

In what follows, we will explicitly check that $\Omega_0$ has the form (\ref{eq:ConstantLi}) and compute its correction which includes non-linear terms and derivatives of the SYM field-strengths. At order $r^0$ and setting the supergravity fields to zero, one can compute $\bar\partial b$ using the equations of motion and (\ref{eq:bghost}), the result is the sum of the following three contributions    
\begin{eqnarray}
\bar\partial (- s^{\alpha} \nabla \hat\la_{\alpha} - \qua \Pi^A T_{A ab} (s \gamma^{ab} \hat\la) - \omega_{\alpha} \Pi^{\alpha})|_{\rm{SYM}}=  \hspace{20mm}  \label{eq:delb1}\\ 
  \omega_{\alpha} \Pi^a J_I  \nabla_a W^{I \alpha}  - J_K f^K_{\esp I J} \,  \omega_{\alpha} W^{I \alpha} d_{\beta} W^{J \beta}  
-  \half J_K f^K_{\esp IJ} \, \omega_{\alpha} W^{I \alpha}  \la^{\gamma} \omega_{\beta} U^{J \beta}_{\gamma}   \, , \nn
\end{eqnarray} 
\begin{eqnarray}
\bar\partial( \frac{1}{2 ( \la \hat\la)} (\omega \gamma_a \hat\la) (\la \gamma^a \Pi) )|_{\rm{SYM}} =  \hspace{50mm}  \label{eq:delb2} \\
 \frac{1}{16( \la \hat\la)^2} (\la \gamma^{ab} \hat\la) J_I  U^{I}_{ab}  (\omega \gamma_c \hat\la) (\la \gamma^c \Pi) - \frac{1}{16 (\la \hat\la)} (\omega \gamma_c)^{\alpha} J_I  U^{I}_{ab} (\gamma^{ab})_{\alpha}^{\esp \gamma}  \hat\la_{\gamma} (\la \gamma^c \Pi) \hspace{5mm} \nn \\
+ \frac{1}{2(\la \hat\la)} (\omega \gamma_a \hat\la)(\la \gamma^a)_{\alpha}  (J_K f^{K}_{\esp I J} \, W^{I \alpha}  d_{\beta} W^{J \beta} +\half J_K  f^K_{\esp I J}  \, W^{I \alpha} \la^{\beta} \omega_{\gamma} U^{J \gamma}_{\beta}  - \Pi^a J_I  \nabla_a W^{I\alpha} ) \, ,  \nn
\end{eqnarray}
and finally
\begin{eqnarray}
\bar\partial( \Pi^a \bar\Gamma_a)|_{{\rm{SYM}}, r^{0}} = - \frac{1}{16 (\la \hat\la)^2} (\la \gamma^{ab} \hat\la) J_I U^I_{ab}  \, \Pi^c (D \gamma_c \hat\la) \hspace{12mm} \label{eq:delb3} \\ + \frac{1}{16(\la \hat\la)} \Pi^c (D \gamma_c)^{\alpha} J_I U^I_{ab} (\gamma^{ab})_{\alpha}^{\esp \beta} \hat\la_{\beta} 
+ \frac{1}{4 (\la \hat\la)} \Pi^a (\hat\la \gamma_a)^{\alpha} \la^{\beta} \omega_{\gamma} J_I \nabla_{\alpha} U^{I \gamma}_{\beta}  \nn \\ - \frac{1}{4 (\la \hat\la)} (D \gamma^a \hat\la) \la^{\alpha} \omega_{\beta} J_I \nabla_a U^{I \beta}_{\alpha} + \frac{1}{2 (\la \hat\la)} (D \gamma^a \hat\la) J_I \nabla_a W^{I \alpha} d_{\alpha} \, . \hspace{5mm} \nn
\end{eqnarray}

The next step is to compute $\delta_B \Omega_0$ with $\Omega_0$ given in (\ref{eq:ConstantLi}) at order $r^0$ in the limit we are considering. Using the BRST transformations of the subsection \ref{BRST} and the ones given in (\ref{eq:BRSTnon}), we have 
\begin{eqnarray}
\delta_B ( \frac{1}{16 (\la \hat\la)^2} ( D \gamma^a \hat\la) (D \gamma^b \hat\la) J_I U^I_{ab} )|_{{\rm{SYM}},r^0} = \hspace{22mm} \label{eq:BRST1} \\
 \frac{1}{8 (\la \hat\la)^2} \Pi_c ( \la \gamma^c \gamma^a \hat\la) (D \gamma^b \hat\la) J_I U^I_{ab} + \frac{1}{16 (\la \hat\la)^2} ( D \gamma^a \hat\la) (D \gamma^b \hat\la)  \la^{\alpha} J_I \nabla_{\alpha} U^I_{ab} \, , \nn 
\end{eqnarray}
and 
\begin{eqnarray}
\delta_B (- \frac{1}{128 ( \la \hat\la)^2} \hat\la_{\alpha} (\gamma^{ab})_{\beta}^{\esp \alpha} \, J_I  U^I_{ab}  \, \, ^{\beta}(\gamma_{cef} \hat\la) (\la \gamma^{ef} \omega) \Pi^c )|_{{\rm{SYM}},r^0} = \label{eq:BRST2} \\
- \frac{1}{128 ( \la \hat\la)^2} \hat\la_{\alpha} (\gamma^{ab})_{\beta}^{\esp \alpha} \, J_I  U^I_{ab}  \, \, ^{\beta}(\gamma_{cef} \hat\la) (\la \gamma^{ef} D) \Pi^c \hspace{10mm}  \nn \\
+ \frac{1}{128 ( \la \hat\la)^2} \hat\la_{\alpha} (\gamma^{ab})_{\beta}^{\esp \alpha} \, J_I  U^I_{ab}  \, \, ^{\beta}(\gamma_{cef} \hat\la) (\la \gamma^{ef} \omega) (\la \gamma^c \Pi)   \nn \hspace{8mm} \\
- \frac{1}{128 ( \la \hat\la)^2} \hat\la_{\alpha} (\gamma^{ab})_{\beta}^{\esp \alpha} \,   \la^{\gamma} J_I \nabla_{\gamma} U^I_{ab}  \, \, ^{\beta}(\gamma_{cef} \hat\la) (\la \gamma^{ef} \omega) \Pi^c  \, .  \nn \hspace{5mm}
\end{eqnarray}
Note that the terms above not having derivatives precise reproduce the terms without derivatives and linear in the SYM fields of $\bar\partial b$ computed previously. To see this, one uses the identity   
\qe
 \frac{1}{16} (\gamma^{ab})_{\alpha}^{\esp \gamma} (\gamma_{ab})_{\beta}^{\esp \delta} = \frac{1}{4} \gamma^a_{\alpha \beta} \gamma_{a}^{\gamma \delta} - \frac{1}{8} \delta^{\gamma}_{\alpha} \delta^{\delta}_{\beta} -\half \delta^{\delta}_{\alpha} \delta^{\gamma}_{\beta}  \, , \label{eq:ImportantIdentity}
\ee
the properties of the gamma matrices and the pure spinor condition. 

To reproduce all the terms in $\bar\partial b$ containing derivatives of the SYM superfields, we need to add terms to $\Omega_0$. 
We will make an ansatz for these new terms and then verify that they are the correct ones. The ansatz is       
\qe
\Omega_1 = \frac{1}{2 (\la \hat\la)} (D \gamma^a \hat\la) J_I \nabla_a W^{I \alpha} \omega_{\alpha}- \frac{1}{4 ( \la \hat\la)^2} (D \gamma^a \hat\la) J_I \nabla_a W^{I \alpha} (\gamma_b)_{\alpha \beta} \la^{\beta} (\omega \gamma^b \hat\la) \, . \label{eq:deltaOmega0}
\ee
Notice that $\Omega_1$ is gauge invariant under the gauge transformation of $\omega_{\alpha}$ of (\ref{eq:gaugeomega}). This property of $\Omega_1$ becomes manifest when 
$\Omega_1$ is rewritten using (\ref{eq:ImportantIdentity}) as    
\qe
\Omega_1 =- \frac{1}{16 (\la \hat\la)^2} (D \gamma^a \hat\la) J_I \nabla_a W^{I \alpha} (\gamma_{bc})_{\alpha}^{\esp \beta} \hat\la_{\beta} (\la \gamma^{bc} \omega) - \frac{1}{8 (\la \hat\la)^2} (D \gamma^a \hat\la) J_I \nabla_a W^{I \alpha} \hat\la_{\alpha} (\la \omega)  \, . \nn 
\ee

The BRST transformations of the two terms of $\Omega_1$ are 
\begin{eqnarray}
\delta_B (\frac{1}{2 (\la \hat\la)} (D \gamma^a \hat\la) J_I \nabla_a W^{I \alpha} \omega_{\alpha})|_{{\rm{SYM}},r^0} = \hspace{22mm}  \label{eq:BRST3} \\
\frac{1}{2 (\la \hat\la)} \Pi_b (\la \gamma^b \gamma^a \hat\la) J_I \nabla_a W^{I \alpha} \omega_{\alpha} -\frac{1}{2 (\la \hat\la)} (D \gamma^a \hat\la)  \la^{\gamma} J_I \nabla_{\gamma} \nabla_a W^{I \alpha} \omega_{\alpha}  \nn \\ 
+ \frac{1}{2 (\la \hat\la)}  (D \gamma^a \hat\la) J_I \nabla_a W^{I \alpha} d_{\alpha} \, ,  \hspace{30mm} \nn 
\end{eqnarray}
and 
\begin{eqnarray}
\delta_B(- \frac{1}{4 ( \la \hat\la)^2} (D \gamma^a \hat\la) J_I \nabla_a W^{I \alpha} (\gamma_b)_{\alpha \beta} \la^{\beta} (\omega \gamma^b \hat\la))|_{{\rm{SYM}},r^0} = \label{eq:BRST4} \\
- \frac{1}{4 ( \la \hat\la)^2} \Pi_c (\la \gamma^c \gamma^a \hat\la) J_I \nabla_a W^{I \alpha} (\gamma_b)_{\alpha \beta} \la^{\beta} (\omega \gamma^b \hat\la) \hspace{10mm}  \nn \\
- \frac{1}{4 ( \la \hat\la)^2} (D \gamma^a \hat\la) J_I \nabla_a W^{I \alpha} (\gamma_b)_{\alpha \beta} \la^{\beta} (D \gamma^b \hat\la) \hspace{13mm} \nn \\
+ \frac{1}{4 ( \la \hat\la)^2} (D \gamma^a \hat\la)  \la^{\gamma} J_I \nabla_{\gamma} \nabla_a W^{I \alpha} (\gamma_b)_{\alpha \beta} \la^{\beta} (\omega \gamma^b \hat\la) \, .  \hspace{7mm}  \nn 
\end{eqnarray}

To see that in fact all the derivative terms in $\bar\partial b$ are reproduced requires several manipulations, the use of the superfields constraints, the Bianchi identities and the following property of the covariant derivatives    
\qe
[ \nabla_A \, , \nabla_B ]_{\pm} W^{I \alpha} = - T_{AB}^{\esp \esp C} \nabla_C W^{I \alpha} + R_{AB\beta}^{\esp \esp \esp \esum \alpha} W^{I \beta} - f^I_{\es JK}  F^K_{AB}  W^{J \alpha} \, , \label{eq:CommutationDerivativesW}
\ee
where $\pm$ means commutator or anticommutator depending on the grading of $A$ and $B$. Firstly, note that the last term of (\ref{eq:BRST3}) is equal to the last term of (\ref{eq:delb3}). Secondly, using the constraint $\half U^{I \alpha}_{\beta} = \nabla_{\beta}W^{I \alpha}$ and (\ref{eq:CommutationDerivativesW}) with the appropriate indices, we can rewrite the third term on the right-hand side of (\ref{eq:delb3}) as   
\qe
\frac{1}{4 (\la \hat\la)} \Pi^a (\hat\la \gamma_a)^{\alpha} \la^{\beta} \omega_{\gamma} J_I \nabla_{\alpha} U^{I \gamma}_{\beta} =  \frac{1}{2 (\la \hat\la)} \Pi^a (\hat\la \gamma_a)^{\alpha} \la^{\beta} \omega_{\gamma} J_I \nabla_{\alpha} \nabla_{\beta} W^{I \gamma} =  \hspace{10mm}  \label{eq:Inter1} \\
- \frac{1}{4 (\la \hat\la)} \Pi^a (\hat\la \gamma_a)^{\alpha} \la^{\beta} \omega_{\gamma} J_I \nabla_{\beta} U^{I \gamma}_{\alpha} - 
 \omega_{\alpha} \Pi^a J_I  \nabla_a W^{I \alpha}  
+\frac{1}{2 (\la \hat\la)} \Pi_b (\la \gamma^b \gamma^a \hat\la) J_I \nabla_a W^{I \alpha} \omega_{\alpha} \, . \nn
\ee
It is immediately to see that the second term in the second line of the expression above cancels the first term on the right-hand side of (\ref{eq:delb1}) and the third term is equal to the first term on the right-hand side of (\ref{eq:BRST3}). Similarly, one has that the fourth term in (\ref{eq:delb3}) takes the form 
\qe
- \frac{1}{4 (\la \hat\la)} (D \gamma^a \hat\la) \la^{\alpha} \omega_{\beta} J_I \nabla_a U^{I \beta}_{\alpha} = - \frac{1}{2 (\la \hat\la)} (D \gamma^a \hat\la) \la^{\alpha} \omega_{\beta} J_I \nabla_a \nabla_{\alpha} W^{I \beta} = \label{eq:Inter2} \\
 -\frac{1}{2 (\la \hat\la)} (D \gamma^a \hat\la)  \la^{\gamma} J_I \nabla_{\gamma} \nabla_a W^{I \alpha} \omega_{\alpha}   + \frac{1}{2 (\la \hat\la)} (D \gamma^a \hat\la)  \la^{\gamma} J_I f^{I}_{\esp JK}  F_{a \gamma}^K W^{J \alpha} \omega_{\alpha} \, . \nn 
\ee
Note that the first term in the second line of the expression above is equal to the second term on the right-hand side of (\ref{eq:BRST3}). To finish the proof that all the terms with derivatives in $\bar\partial b$ are reproduced,  we need to manipulate some of the terms coming from the BRST variations as well. Using the identity (\ref{eq:ImportantIdentity}), the pure spinor condition, $F_{\alpha \beta}=0$ and  (\ref{eq:CommutationDerivativesW}), one can show that the last term of (\ref{eq:BRST2}) takes the form 
\qe
- \frac{1}{128 ( \la \hat\la)^2} \hat\la_{\alpha} (\gamma^{ab})_{\beta}^{\esp \alpha} \, \la^{\gamma}  J_I  \nabla_{\gamma} U^I_{ab}  \, \, ^{\beta}(\gamma_{cef} \hat\la) (\la \gamma^{ef} \omega) \Pi^c = \,    \hspace{10mm} \label{eq:Inter3} \\
- \frac{1}{4 (\la \hat\la)} \Pi^a (\hat\la \gamma_a)^{\alpha} \la^{\beta} \omega_{\gamma} J_I \nabla_{\beta} U^{I \gamma}_{\alpha} - \frac{1}{8 (\la \hat\la)^2} \Pi^{a} (\la \gamma^b \gamma_a \hat\la) (\omega \gamma^c \hat\la) \la^{\alpha} J_I \nabla_{\alpha} U^{I}_{c b} \, .  \nn
\ee
One can verify that the first term on the right-hand side of the expression above is equal to the first term in the second line of (\ref{eq:Inter1}). 
In addition, using the following identities valid in the limit we are considering (the second one follows from $\half U^{I \beta}_{\alpha} = \nabla_{\alpha} W^{I \beta}$ and the pure spinor condition)  
\qe
(\gamma^a)_{\alpha \beta} \nabla_a W^{I \gamma}= -(\nabla_{\alpha} \nabla_{\beta} + \nabla_{\beta} \nabla_{\alpha}) W^{I \gamma} \, , \esp \esp  \la^{\alpha} (\la \gamma_a)_{\beta}  (\nabla_{\alpha} W^{I \beta}) = 0 \, ,
\ee
one can show that the first term on the right-hand side of (\ref{eq:BRST4}) is  
\qe
- \frac{1}{4 ( \la \hat\la)^2} \Pi_c (\la \gamma^c \gamma^a \hat\la) J_I \nabla_a W^{I \alpha} (\gamma_b)_{\alpha \beta} \la^{\beta} (\omega \gamma^b \hat\la) = \hspace{30mm}  \nn \\
 \frac{1}{4 ( \la \hat\la)^2} \Pi_c (\la \gamma^a \gamma^c \hat\la) J_I \nabla_a W^{I \alpha} (\gamma_b)_{\alpha \beta} \la^{\beta} (\omega \gamma^b \hat\la) - \frac{1}{2 ( \la \hat\la)} \Pi^a  J_I \nabla_a W^{I \alpha} (\gamma_b)_{\alpha \beta} \la^{\beta} (\omega \gamma^b \hat\la) = \nn \\
\frac{1}{8 (\la \hat\la)^2} \Pi^{a} (\la \gamma^b \gamma_a \hat\la) (\omega \gamma^c \hat\la) \la^{\alpha} J_I \nabla_{\alpha} U^{I}_{c b} - \frac{1}{2 ( \la \hat\la)} \Pi^a  J_I \nabla_a W^{I \alpha} (\gamma_b)_{\alpha \beta} \la^{\beta} (\omega \gamma^b \hat\la) \, . \nn \hspace{8mm}
\ee
Note that the last term above is equal to the last term in (\ref{eq:delb2}) and the first term in the last line above cancels the last term of (\ref{eq:Inter3}). In addition, using the Bianchi identity (\ref{eq:BianchiF}) in the limit where the supergravity superfields are zero, it is possible to show that the last term in (\ref{eq:BRST1}) cancels the second term on the right-hand side of (\ref{eq:BRST4}).
Finally, one can show that the last term in (\ref{eq:BRST4}) is equal to    
\qe
 \frac{1}{4 ( \la \hat\la)^2} (D \gamma^a \hat\la)  \la^{\gamma} J_I \nabla_{\gamma} \nabla_a W^{I \alpha} (\gamma_b)_{\alpha \beta} \la^{\beta} (\omega \gamma^b \hat\la)
=  \label{eq:Inter4}  \\ 
- \frac{1}{4 ( \la \hat\la)^2} (D \gamma^a \hat\la)  \la^{\gamma} J_I f^I_{\esp JK} F^K_{\gamma a} W^{J \alpha} (\gamma_b)_{\alpha \beta} \la^{\beta} (\omega \gamma^b \hat\la) \, . \nn 
\ee
This concludes the proof that all terms linear in the SYM superfields containing derivatives or not in $\bar\partial b$ are reproduced by $\delta_B (\Omega_0 +  \Omega_1)$ at order $r^0$ as only non-linear terms are left. The non-linear terms in $\bar\partial b$ come from (\ref{eq:delb1}), (\ref{eq:delb2}) and (\ref{eq:Inter2}), they are
\qe
- J_K f^K_{\esp I J} \,  \omega_{\alpha} W^{I \alpha} d_{\beta} W^{J \beta}  
-  \half J_K f^K_{\esp IJ} \, \omega_{\alpha} W^{I \alpha}  \la^{\gamma} \omega_{\beta} U^{J \beta}_{\gamma}  \hspace{15mm}  \nn \\
+ \frac{1}{2(\la \hat\la)} (\omega \gamma_a \hat\la)(\la \gamma^a)_{\alpha}  (J_K f^{K}_{\esp I J} \, W^{I \alpha}  d_{\beta} W^{J \beta} +\half J_K  f^K_{\esp I J}  \, W^{I \alpha} \la^{\beta} \omega_{\gamma} U^{J \gamma}_{\beta} )   \nn \\
+\frac{1}{2 (\la \hat\la)} (D \gamma^a \hat\la)  \la^{\gamma} J_I f^{I}_{\esp JK}  F_{a \gamma}^K W^{J \alpha} \omega_{\alpha}  \, . \nn \hspace{27mm}
 \ee
The only non-linear term from the previous BRST variations is the last one in (\ref{eq:Inter4}). It is not difficult to see, using the BRST transformations of both the subsection \ref{BRST} and (\ref{eq:BRSTnon}) and the constraints  (\ref{eq:Constraints}), that $\delta_B \Omega_2$ with   
\qe
\Omega_2 = \frac{1}{2} J_{K} f^{K}_{\esp I J} \, \omega_{\alpha} W^{I \alpha}  \omega_{\beta} W^{J \beta} - \frac{1}{2 ( \la \hat\la)} J_{K} f^{K}_{\esp I J} \,  (\omega \gamma^a \hat\la)( \la \gamma_a)_{\alpha} W^{I \alpha} \omega_{\beta} W^{J \beta} \label{eq:Omega1} \\
+ \frac{1}{8 (\la \hat\la)^2} J_{K} f^{K}_{\esp I J} \, W^{I \alpha} (\gamma_a)_{\alpha \beta} \la^{\beta} (\omega \gamma^a \hat\la) W^{J \gamma} (\gamma_b)_{\gamma \delta} \la^{\delta} (\omega \gamma^b \hat\la) \, ,  \nn \hspace{10mm} 
\ee
reproduces all the non-linear terms in $\bar\partial b$ and cancels the non-linear term coming from the previous BRST variations. Note that $\Omega_2$ can be rewritten as manifestly gauge invariant as          
\begin{eqnarray}
 \Omega_2 = \frac{1}{2 (\la \hat\la)^2} J_K f^{K}_{\esp I J} \, [\,  \frac{1}{8} W^{I \alpha} (\gamma_{ab})_{\alpha}^{\esp \beta} \hat\la_{\beta} (\la \gamma^{ab} \omega)+ \frac{1}{4} W^{I \alpha} \hat\la_{\alpha} (\la \omega) ]  \nn \\ \nn \\  \times [\,  \frac{1}{8} W^{J \gamma} (\gamma_{cd})_{\gamma}^{\esp \delta} \hat\la_{\delta} (\la \gamma^{cd} \omega)+ \frac{1}{4} W^{J \gamma} \hat\la_{\gamma} (\la \omega) ] \, , \nn \hspace{15mm} 
\end{eqnarray}
where we have used (\ref{eq:ImportantIdentity}). 

This concludes the construction of $\Omega$ at order $r^0$ in a super-Yang-Mills heterotic background when all the background supergravity superfields are set to zero.
The next step is to construct $\Omega$ at order $r$ in the same limit. To achieve this, one first compute $\bar\partial b$ at order $r$ using the equations of motion. It is possible to show that all the linear terms without derivatives are reproduce by $\delta_B \Omega_0$ where $\Omega_0$ is given in (\ref{eq:ConstantLi}). One can show that the terms from $\delta_B \Omega_0$ containing derivatives  together with both the terms of order $r$ of  $\delta_B \Omega_1$ with $\Omega_1$ given in (\ref{eq:deltaOmega0}) and $\delta_B \Omega_3$ with 
\qe
\Omega_3  = -\frac{1}{16 (\la \hat\la)^2} [ J_I  \nabla_c W^{I \alpha} \omega_{\alpha} - \frac{1}{2 (\la \hat\la)}  J_I  \nabla_c W^{I \alpha}(\gamma_d)_{\alpha \beta}  \la^{\beta} (\omega \gamma^d \hat\la) ] (r \gamma^{c}_{\esp a b} \hat\la)(\la \gamma^{ab} \omega) \, ,  \nn 
\ee
reproduce all the linear terms of order $r$ in $\bar\partial b$. The procedure to prove this is similar to the one described at order $r^0$ and the details will be omitted.  
In addition, all the non-linear terms are canceled or reproduced by $\delta_B \Omega_2$. 
At order $r^2$ and $r^3$, one follows the same steps described for lower orders in $r$. It is possible to show that all terms in $\bar\partial b$ at this order are reproduced by $\delta_B ( \Omega_0 + \Omega_3)$, however the details will be again omitted as the procedure is similar to the case $r^0$. This concludes the construction of $\Omega$ in a super-Yang-Mills background.



\subsection{The case where $s^{\alpha}=\hat\omega^{\alpha}=0$}

In this section, the case where both the supergravity and super-Yang-Mills background superfields are non-vanishing will be considered. However, a simplifying assumption will be made, we are going to consider $s^{\alpha}=\hat\omega^{\alpha}=0$. In order to construct $\Omega$, we compute $\bar\partial b$ using the equations of motion. 
Observing the result, one verify after a few manipulations that at any order in $r$ the terms linear in $\overline{\Pi}^A T_{A ab}$ can be combined with the terms linear in $J_I U^I_{ab}$ as a function of the combination $\overline{\Pi}^A T_{A ab} + \half J_I U^I_{ab}$. So, we make the ansatz that $\Omega_0$ of 
(\ref{eq:ConstantLi}) has to be modified to 
\qe
(\Omega_0)_{new}= \frac{1}{4} ( \overline{\Pi}^A T_{A ab} + \half J_I U^I_{ab} ) \, \hat\la_{\beta} (\gamma^{ab})_{\alpha}^{\esp \beta} \, \frac{\partial b}{\partial r_{\alpha}} \, .
\ee
In addition, further analyzing the result of $\bar\partial  b$, we add new terms to $\Omega_1$ and $\Omega_3$, which are  
\qe
\Omega_4 = \overline{\Pi}^b T_{a b}^{\esp \esum \alpha} \omega_{\alpha} \bar\Gamma^{a}
- \frac{1}{2 (\la \hat\la)} \overline{\Pi}^b T_{a b}^{\esp \esum \alpha} (\gamma_c)_{\alpha \beta} \la^{\beta} (\omega \gamma^c \hat\la) \bar\Gamma^a \, ,
\ee
where $\bar\Gamma^a$ are the generalized RNS-like variables defined in (\ref{eq:GeneralizedRNS}). Using the BRST transformations, one can show that indeed all the terms in $\bar\partial b$ with  $s^{\alpha}=\hat\omega^{\alpha}=0$ are reproduced. For the terms containing the super-Yang-Mills fields, one follows the procedure describe in the previous subsection, but this time keeping the terms with supergravity fields. To show that the remaining terms in $\bar\partial b$ are reproduced, one uses (\ref{eq:TheRTwoIndices}) and the following identities. The Bianchi identity (\ref{eq:BianchiT}) with the indices $[ \beta ab ]$ and the constraints (\ref{eq:Constraints}) imply
\qe
\nabla_{\beta} T_{ab}^{\esp \es \alpha} =R_{ab\beta}^{\esp \esp \esp \alpha} - T_{\beta a}^{\esp \es c} \, T_{c b}^{\esp \es \alpha} - T_{b \beta}^{\esp \es c} \, T_{c a}^{\esp \es \alpha} \, .  \nn
\ee
The same Bianchi identity (\ref{eq:BianchiT}), but with the indices $[ \alpha \beta a]$ implies
\qe
(\gamma^b)_{\alpha \beta} T_{b a}^{\esp \es \gamma} = R_{a \alpha \beta}^{\esp \esp \esp \gamma} - R_{\beta a \alpha}^{\esp \esp \esp \gamma} \nn.
\ee
Finally, one can show that
\qe
R_{c \alpha ab}= \nabla_c T_{\alpha ab} + T_{a c}^{\esp \es \beta} (\gamma_b)_{\beta \alpha} + T_{c b}^{\esp \es \beta}(\gamma_a)_{\beta \alpha} \, . \nn
\ee
In order to prove the identity above, we have used that
\qe
R_{c \alpha ab} = \qua ( R_{[c \alpha a] b} - R_{[c \alpha b] a} + R_{[\alpha a b]c}) \, ,
\ee
the Bianchi Identity  (\ref{eq:BianchiT}) three times, the Bianchi Identity (\ref{eq:BianchiH}) and $T_{abc}=-H_{abc}$ as proved in  (\ref{eq:TwithH}).

\subsection{The general case}

In this subsection, we will consider the general case. All the terms in $\bar\partial b$ not containing either $s^{\alpha}$ or $\hat\omega^{\alpha}$ have already been written as  BRST exact in the previous two subsections. Thus, the remaining terms at order $r^0$ in $\bar\partial b$ are   
\begin{eqnarray}
\bar\partial (- s^{\alpha} \nabla \hat\la_{\alpha} - \qua \Pi^A T_{A ab} (s \gamma^{ab} \hat\la) - \omega_{\alpha} \Pi^{\alpha})|_{s, \hat\omega} =  \label{eq:delbsw1} \\ 
\Pi^d \overline{\Pi}^A R_{A d} (s \hat\la) - \qua \overline{\Pi}^A \Pi^d \mathfrak{R}_{d A ab} (s \gamma^{ab} \hat\la)  \, ,    \hspace{7mm}\nn
\end{eqnarray} 
\begin{eqnarray}
\bar\partial( \frac{1}{2 ( \la \hat\la)} (\omega \gamma_a \hat\la) (\la \gamma^a \Pi) )|_{s,\hat\omega} =  \hspace{35mm}   \label{eq:delbsw2} \\
\frac{1}{2 (\la \hat\la)} (\hat\la \gamma_a)^{\alpha} \overline{\Pi}^A R_{A \alpha} (s \hat\la)(\la \gamma^a \Pi) + \frac{1}{8 (\la \hat\la)} (\hat\la \gamma_d)^{\alpha} \overline{\Pi}^c \mathcal{R}_{c \alpha ab} ( s \gamma^{ab} \hat\la) (\la \gamma^{d} \Pi) \, ,  \nn 
\end{eqnarray}
and finally
\begin{eqnarray}
\bar\partial( \Pi^a \bar\Gamma_a)|_{s,\hat\omega,r^0} = - \frac{1}{2 (\la \hat\la)} (D \gamma_a \hat\la) (\bar\partial \Pi^a)|_{s,\hat\omega} - \frac{1}{2 (\la \hat\la)} \Pi^a (\hat\la \gamma_a)^{\alpha} (\bar\partial d_{\alpha})|_{s,\hat\omega}  \hspace{10mm} \label{eq:delbsw3} \\
+ \frac{3 (\la \Omega)}{2 (\la \hat\la)}  \, \Pi^a (\hat\la \gamma_a)^{\alpha} \overline{\Pi}^A R_{A \alpha} (s \hat\la) -\frac{1}{8 (\la \hat\la)} \Pi^c (\hat\la \gamma_c)^{\alpha} \la^{\beta} T_{\beta ab} (\gamma^{ab})_{\alpha}^{\esp \gamma} \overline{\Pi}^A R_{A \gamma} (s \hat\la) \nn \hspace{8mm} \\
+ \frac{3 (\la \Omega)}{8 (\la \hat\la)}  \, \Pi^d (\hat\la \gamma_d)^{\alpha} \overline{\Pi}^c \mathcal{R}_{c \alpha ab} (s \gamma^{ab} \hat\la) 
-\frac{1}{32 (\la \hat\la)} \Pi^c (\hat\la \gamma_c)^{\alpha} \la^{\beta} T_{\beta ab} (\gamma^{ab})_{\alpha}^{\esp \gamma} \overline{\Pi}^d \mathcal{R}_{d \gamma ef} (s \gamma^{ef} \hat\la) \, , \nn 
\end{eqnarray}
where $\bar\partial d_{\alpha}|_{s,\hat\omega} $ and $\bar\partial \Pi^a |_{s,\hat\omega} $ can be read from (\ref{eq:eomd}) and  (\ref{eq:eomPi}) respectively.  Let us consider first the terms involving the curvatures with two indices.  One can show that the terms above from $\bar\partial b$ can be written as $\delta_B \Omega_{0s} |_{r^0}$ with   
\qe
\Omega_{0s}= - \frac{1}{2 (\la \hat\la)} \Pi^a (\hat\la \gamma_a)^{\alpha} \overline{\Pi}^A R_{A \alpha} (s \hat\la) +   \frac{1}{2 (\la \hat\la)} (D \gamma^a \hat\la) \overline{\Pi}^A R_{A a} (s \hat\la) \, .
\ee
The proof is straightforward, it follows from the identities
\qe
\nabla_{[A} R_{BC ]} + T_{[AB}^{\esp \esp \es F} R_{|F| C]} = 0 \, , \quad \quad \overline\nabla \la^{\alpha}= - \half \la^{\beta} J_I U^{I \alpha}_{\beta} \, . \nn 
\ee  
One can compute the terms in $\bar\partial b$ of higher orders in $r$ using the equations of motion. In this case, one has to add to $\Omega_{0s}$ terms that depend on $r$. The additional terms are given in the next section and the proof that they are the correct ones only uses the two identities above.  

Removing the terms with curvatures with two indices from (\ref{eq:delbsw1}), (\ref{eq:delbsw2}) and  (\ref{eq:delbsw3}), we are left with terms with curvatures with four indices and torsions. These remaining terms can also be written as BRST exact. They are equal to $\delta_B \Omega_{1s}|_{r^0}$, where 
\qe
\Omega_{1s}=- \frac{1}{8(\la \hat\la)}  \Pi^c  (\hat\la \gamma_c)^{\alpha} \overline{\Pi}^d \mathcal{R}_{d \alpha ab} (s \gamma^{ab} \hat\la) -\frac{1}{8 (\la \hat\la)} (D \gamma^c \hat\la) \overline{\Pi}^D \mathfrak{R}_{c D ab} (s \gamma^{ab} \hat\la) \, . \nn 
\ee
To prove it, we have used the following two non-trivial identities
\qe
(\gamma^c)_{\alpha \beta} \mathfrak{R}_{c A ab} = \nabla_{\alpha} \mathcal{R}_{A \beta ab} + \nabla_{\beta} \mathcal{R}_{A \alpha ab}  \hspace{15mm}  \label{eq:NonTrivial1} \\ \nn \\
-T_{A \alpha}^{\esp \esp c} \mathcal{R}_{c \beta ab} -T_{A \beta}^{\esp \esp c} \mathcal{R}_{c \alpha ab} - \mathcal{R}_{A \alpha c [ a} T_{b] \beta}^{\esp \esp c}  - \mathcal{R}_{A \beta c [ a} T_{b] \alpha}^{\esp \esp c} \, , \nn 
\ee
and 
\qe
\nabla_{[c} \mathcal{R}_{A] \alpha ab} + T_{c A}^{\esp \esp B} \mathcal{R}_{B \alpha ab} + \mathcal{R}_{A \alpha d [a} T_{b] c}^{\esp \esp d} - (-)^A \mathcal{R}_{c \alpha d [a} T_{b] A}^{\esp \esp d} = \label{eq:NonTrivial2} \\ \nn \\
T_{A \alpha}^{\esp \esp d} \mathfrak{R}_{c d ab} + (-)^A T_{c \alpha}^{\esp \esp d} \mathfrak{R}_{d D ab} - (-)^A \nabla_{\alpha} \mathfrak{R}_{c A ab} - \mathfrak{R}_{c A d [a} T_{b] \alpha}^{\esp \esp d} \, . \nn 
\ee
These identities can be checked after tedious calculations by substituting the definitions of $\mathcal{R}$ and $\mathfrak{R}$ given in  (\ref{eq:DefinitionRT}) and  (\ref{eq:DefinitionRTT}), using the Bianchi Identities (\ref{eq:BianchiT}) and (\ref{eq:BianchiR}) several times and the following relation among covariant derivatives 
\qe
&[ \nabla_A , \nabla_B]_{\pm} T_{CD}^{\esp \esp E} = - T_{AB}^{\esp \esp F} \nabla_F T_{CD}^{\esp \esp E}  \nn \\ \nn \\
&- R_{ABC}^{\esp \esp \esp \es F} T_{FD}^{\esp \esp E} - (-)^{C(D+F)} R_{ABD}^{\esp \esp \esp \es F} T_{C F}^{\esp \esp E} + (-)^{(A+B)(F+C+D)} T_{CD}^{\esp \esp F} R_{ABF}^{\esp \esp \esp \es E} \, , \nn 
\ee
where $\pm$ means commutator or anticommutator depending on $A$ and $B$. 

This completes the construction of $\Omega$ at order $r^0$. Using the equations of motion to evaluate $\bar\partial b$, it is easy to find the terms with $s^{\alpha}$ and $\hat\omega^{\alpha}$ which depends on $\mathcal{R}$ and $\mathfrak{R}$ and are higher order in $r$.  
These terms are again BRST exact if one modifies $\Omega_{1s}$ by terms depending on $r$. These additional terms of $\Omega_{1s}$ are given in the next section and the proof that they are the correct ones is straightforward and follows from the equalities (\ref{eq:NonTrivial1}) and (\ref{eq:NonTrivial2}).

\section{The Complete $\Omega$}

In this section, the complete expression for $\Omega$ is presented. The $\Omega$ is a solution of the equation $\bar\partial b = \delta_B \Omega$ and it is defined up to a BRST exact term. Our $\Omega$ is  

\qe
\Omega  = \frac{1}{2} J_{K} f^{K}_{\esp I J} \, \omega_{\alpha} W^{I \alpha}    \omega_{\beta} W^{J \beta} - \frac{1}{2 ( \la \hat\la)} J_{K} f^{K}_{\esp I J} \,  W^{I \alpha} (\gamma_a)_{\alpha \beta} \la^{\beta} (\omega \gamma^a \hat\la) \omega_{\gamma} W^{J \gamma} \hspace{10mm} \nn \\ \nn \\ 
+ \frac{1}{8 (\la \hat\la)^2} J_{K} f^{K}_{\esp I J} \, W^{I \alpha} (\gamma_a)_{\alpha \beta} \la^{\beta} (\omega \gamma^a \hat\la) W^{J \gamma} (\gamma_b)_{\gamma \delta} \la^{\delta} (\omega \gamma^b \hat\la) \hspace{25mm} \nn \\ \nn \\
+\frac{1}{4} ( \overline{\Pi}^A T_{A ab} + \half J_I U^I_{ab} ) \, \hat\la_{\beta} (\gamma^{ab})_{\alpha}^{\esp \beta} \, \frac{\partial b}{\partial r_{\alpha}} \hspace{40mm} \nn \\ \nn \\
+( J_I \nabla_a W^{I \alpha} + \overline{\Pi}^b T_{a b}^{\esp \esum \alpha}) \omega_{\alpha} \bar\Gamma^{a}
- \frac{1}{2 (\la \hat\la)} (J_I \nabla_a W^{I \alpha} + \overline{\Pi}^b T_{a b}^{\esp \esum \alpha}) (\gamma_c)_{\alpha \beta} \la^{\beta} (\omega \gamma^c \hat\la) \bar\Gamma^a \nn \hspace{5mm}  \\ \nn \hspace{40mm}  \\
- \frac{1}{2 (\la \hat\la)}  \Pi^a  (\hat\la \gamma_a)^{\alpha}  \overline{\Pi}^A R_{A \alpha} (s \hat\la)+ \overline{\Pi}^A R_{A a} (s \hat\la) \bar\Gamma^a
- \frac{ 1}{ 4 (\la \hat\la)^2 } (\la \gamma^{ab} r)(\hat\la \gamma_a)^{\alpha} \overline{\Pi}^A R_{A \alpha} (s \hat\la)   \bar\Gamma_b  \nn \\ \nn \\
- \frac{1}{8(\la \hat\la)} \Pi^c (\hat\la \gamma_{c})^{\alpha} \overline{\Pi}^d \mathcal{R}_{d \alpha ab} (s \gamma^{ab} \hat\la)   - \frac{1}{4} \overline{\Pi}^D \mathfrak{R}_{c D ab} (s \gamma^{ab} \hat\la) \bar\Gamma^c \nn \hspace{23mm}  \\ \nn \\
- \frac{1}{16 (\la \hat\la)^2} (\la \gamma^{ab} r) (\hat\la \gamma_a)^{\alpha} \overline{\Pi}^D \mathcal{R}_{D \alpha c d} (s \gamma^{cd} \hat\la) \bar\Gamma_b \, . \hspace{30mm} \nn 
\ee

\section{Acknowledgements}

We would like to thank Thales Agricola, Nathan Berkovits and Osvaldo Chandia for very useful comments and discussions. We would like to thank the warm hospitality of the Perimeter Institute where this work was finished. We would like to thank FAPESP grants 2013/12416-7 and 2015/01135-2 for financial support.

\appendix
\section{Notations and Conventions} 

In this paper, we mainly follow the conventions of Wess and Bagger \cite{WessBaggerLivro} and of  Chandia and Vallilo \cite{ConventionsCV}. The curved superspace indices are denoted by $M$ and take the values 
$M=(m, \mu)$ with $m=0, \ldots ,9$ and  $\mu=1, \ldots, 16$. The tangent superspace  indices are denoted by $A=(a, \alpha)$, with $a=0, \ldots ,9$ and $\alpha = 1, \ldots, 16$.   
Being $Z^M$ the $\mathcal{N}=1$ $D=10$ curved superspace coordinates, one has
\qe
Z^M Z^N = (-)^{MN} Z^N Z^M \, . \nn 
\ee
In order to define differential forms, one introduces the exterior product, which obeys   

\qe
d Z^M \wedge \, dZ^N = - (-)^{MN} \, dZ^N \wedge \, dZ^M  \, . \nonumber
\ee
A  $p$-form $\Theta$ is defined by 
\qe
\Theta = \frac{1}{p!} \, dZ^{M_1} \wedge \, dZ^{M_2} \wedge  \, \ldots \,  \wedge \, dZ^{M_p} \, \Theta_{M_p \ldots M_1} \, . \nonumber  
\ee

The exterior derivative $d$ is an operator which maps $p$-forms into $p+1$-forms, its action on $\Theta$ is    
\qe
d \Theta =  \frac{1}{p!} \, dZ^{M_1} \wedge \, dZ^{M_2} \wedge \, \ldots \,  \wedge \,  dZ^{M_p} \wedge \,  dZ^N \, \frac{\partial}{\partial Z^N} \Theta_{M_p \ldots M_1} \, . \nn 
\ee

An important property of the exterior derivative that follows from the above definitions is 
\qe
d (\Theta \Sigma) = \Theta (d \Sigma) + (-)^q (d \Theta) \Sigma \, ,  \nn 
\ee
where $\Sigma$ is a $q$-form. 

\subsection{Connections and Covariant Derivatives}

Three different connections appear in this paper: the gauge connection, the Lorentz connection and the scaling connection. The one-form gauge connection taking values in a Lie Algebra is 
\qe
A = dZ^M A_{M}^{ I} T_I \, , \nn
\ee
where $T_I$ are the Lie Algebra generators. The one-form Lorentz connection and scaling connection are denoted respectively by   
\qe
\Omega_{a}^{\esp b}  = dZ^M \Omega_{M a}^{\esp \esp \esp  b} \, ,  \quad  \quad  \hat \Omega_s = dZ^M \Omega_M \, , \nn 
\ee
with $\Omega_{M a b} = -\Omega_{M b a}$.  These two connections appear in the decomposition of $\Omega_{M \alpha}^{\esp \esp \es \beta}$ as
\qe
\Omega_{M \alpha}^{\esp \esp \es \beta} = \Omega_{M} \, \delta_{\alpha}^{\beta} + \frac{1}{4} \Omega_{M ab} (\gamma^{ab})_{\alpha}^{\esp \beta} \, . \nn
\ee

Using the one-form connections given above, one can define covariant derivatives of forms. Consider a $p$-form $\Sigma_{A}^{\esp  B}=\Sigma^{I B}_{ A} \, T_I$ taking values in a Lie Algebra. The covariant derivative of $\Sigma_{A}^{\esp B}$ is defined by   
\qe
\na \Sigma_{A}^{\, \, \, \,  B} = d \Sigma_{A}^{\, \, \, \, B}  - \Omega_{A}^{\, \, \, \, C} \wedge \Sigma_{C}^{\esp B} + \Sigma_{A}^{\esp C} \wedge \, \Omega_C^{\esp B}- \Sigma_{A}^{\esp B}  \wedge \, A + (-)^p A \wedge \, \Sigma_{A}^{\esp B} \, . \nn 
\ee  
Using the previous definitions and 
\qe
\na \Sigma_{A}^{\, \, \, \,  B} =  \frac{1}{p!} \, dZ^{M_1} \wedge \, dZ^{M_2} \wedge \, \ldots \,  \wedge \,  dZ^{M_p} \wedge \,  dZ^N \, \na_N (\Sigma^{I}_{M_p \ldots M_1})_{A}^{\, \, \, \,  B} T_I \, , \nn 
\ee
it is possible to deduce an expression for the covariant derivative of the components of the form $\Sigma_A^{\esp B}$. 

\subsection{Supervielbein, Torsions and Curvatures}

The one-form supervielbein is defined by 
\qe
E^A = dZ^M E_{M}^{\esp \es A} \, .  \nn 
\ee
The inverse supervielbein matrix is defined implicitly by the relations 
\qe
E_{M}^{\esp \es A} E_{A}^{\esp N} = \delta_M^N \, , \quad  \quad \quad E_{A}^{\esp  M} E_{M}^{\esp \es B} = \delta_A^B \, . \nn 
\ee

Using the definitions of the one-form supervielbein, the one-form connections and the covariant derivative, one defines the two-form torsion $T^A$  as

\qe
T^{A} = \na E^A = d E^A + E^B \hspace{-1mm} \wedge \Omega_B^{\esp A} = \frac{1}{2} dZ^M \hspace{-1.5mm} \wedge dZ^N T_{NM}^{\esp \esp \es A} = \frac{1}{2} E^C \hspace{-1mm}  \wedge E^B T_{BC}^{\esp \esp A}\, ,  \nn
\ee
or in components,
\qe
T_{NM}^{\esp \esp \es A} = \partial_{[N} E_{M]}^{\esp \es A} + (-)^{N(B+M)} E_M^{\esp \es B} \Omega_{N B}^{\esp \esp \es A} - (-)^{MB} E_N^{\esp \es B}  \Omega_{MB}^{\esp \esp \es A} \, , \label{eq:TorsionDef}
\ee 
where the brackets means antisymmetrization of the indices without any additional numerical factor, i.e. $\partial_{[N} E_{M]}^{\esp \es A}=\partial_{N} E_{M}^{\esp \es A} -(-)^{MN}\partial_{M} E_{N}^{\esp \es A}$.  The curvature two-form $R_A^{\esp B}$ is given by 
\qe
R_A^{\esp B} = d \Omega_A^{\esp B}  + \Omega_A^{\esp C} \hspace{-1mm}  \wedge \Omega_{C}^{\esp B} = \frac{1}{2} dZ^M \hspace{-1mm}  \wedge dZ^N R_{NMA}^{\esp \esp \esp \esp  B} = \frac{1}{2} E^C \hspace{-1mm}  \wedge E^D R_{DCA}^{\esp \esp \esp \esp B} \, ,  \nn
\ee
or in components
\qe
R_{NMA}^{\esp \esp \esp \esp B} = \partial_{[N} \Omega_{M] A}^{\esp \esp \esp  \es B}  + (-)^{N(M+A+C)} \Omega_{M A}^{\esp \esp \esp C} \Omega_{N C}^{\esp \esp \es B} 
-(-)^{M(A+C)} \Omega_{NA}^{\esp \esp \es C} \Omega_{M C}^{\esp \esp \es B} \, . \nn 
\ee
Note that due to the fact that $\Omega_{M \alpha}^{\esp \esp \es \beta}$ has the decomposition (\ref{eq:DecompositionOU}), it is possible to show that
\qe
R_{NM \alpha}^{\esp \esp \esp \esp \beta} = R_{NM} \delta_{\alpha}^{\beta} + \frac{1}{4} R_{NMab} (\gamma^{ab})_{\alpha}^{\esp  \beta}  \, , \nn
\ee
where the curvature with two indices is constructed from the scaling connection.

\subsection{The super-Yang-Mills field-strength and the three-form field-strength}

The two-form super-Yang-Mills field-strength is constructed from the gauge connection as 
\qe
F = F^I T_I = d A - A  \wedge A = \frac{1}{2} dZ^M \hspace{-1mm}  \wedge dZ^N F_{NM} = \frac{1}{2} E^A \hspace{-1mm}  \wedge E^B F_{BA} \, , \nn
\label{eq:Fdef}
\ee
or in components, 
\qe
F_{NM}^I = \partial_{[N} A^I_{M]} - (-)^{NM} A^J_{M} A^K_{N} f^{I}_{\esp JK} \, , \nn 
\ee
where $f^I_{\esp JK} $ are the structure constants of a Lie Algebra, $[ \, T_J \, , \, T_K \, ] = f^{I}_{\esp \,  JK} T_I  $.  The three-form field-strength $H$ is written in terms  of the two-form superfield potential $B$ as
\qe
H = dB = \frac{1}{6} dZ^M \hspace{-1mm}  \wedge dZ^N \hspace{-1mm}  \wedge dZ^P H_{PNM} = \frac{1}{6} E^C \hspace{-1mm}  \wedge E^B \hspace{-1mm}  \wedge E^A H_{ABC} \, ,  \nn 
\ee
or in components 
\qe
H_{PNM} = \frac{1}{2} \partial_{[P} B_{NM]} \, , \nn
\ee
where again the brackets mean antisymmetrization of the indices with no additional numerical factors. 

\subsection{Bianchi Identities} 

In this subsection, we collect all the Bianchi identities. Using the definitions of the previous subsections of this Appendix, we have 
\qe
T^D = \nabla E^D \, , \quad \nabla T^D = E^B \hspace{-1mm}  \wedge R_B^{\esp D}  \es \esum \Rightarrow \es \esum \nabla_{[A} T_{BC]}^{\esp \esp \es D} - R_{[ABC]}^{\esp \esp \esp \esp D} + T_{[AB}^{ \esp \esp \es F} T_{|F| C]}^{\esp \esp \esp D} =0 \, ,
\label{eq:BianchiT}
\\ \nn \\
 \nabla F = 0 \quad \Rightarrow \quad \nabla_{[A} F^I_{BC]} + T_{[AB}^{\esp \esp \es D} F^I_{|D| C]} =0 \, , \hspace{30mm}
\label{eq:BianchiF}
\\ \nn \\
 \nabla R_{D}^{\esp E} =0 \quad \Rightarrow \quad \nabla_{[A} R_{BC]D}^{\esp \esp \esp \esp E} + T_{[AB}^{\esp \esp \es F} R_{|F| C] D}^{ \esp \esp \esp \esp \es \, E} =0 \, ,  
\label{eq:BianchiR} \hspace{25mm}  \\ \nn \\ 
 d H =0 \quad \Rightarrow \quad \nabla_{[A} H_{BCD]} + \frac{3}{2} T_{[A B}^{\esp \esp \es E} H_{|E| CD]} =0 \, ,
\label{eq:BianchiH} \hspace{27mm}
\ee
where the indices inside the brackets but outside the double bars 
{\small $|\,  |$} are antisymmetrized.

\end{document}